# A Proposed Exact Integer Value for Avogadro's Number
by
Ronald F. Fox and Theodore P. Hill
12/8/2006

From the early 1600's until the latter part of the twentieth century, the speed of light was determined experimentally by various methods, based on the physical definition of a second at the time of the experiment (e.g., in recent times, via the integer number of cycles in the Cesium-133 standard, i.e. 9,192,631,770 hertz vibrations of a cesium atom), and an artifact meter stick. On October 21, 1983, the roles of the two constants were reversed when the speed of light *c* was *defined* as *exactly* 299,792,458 meters per second, and the distance 1m has since been approximated experimentally using these values of the speed of light and the second. This new numerical value chosen for *c* was within the best known limits of accuracy at the time, and eliminated the necessity of the artifact meter stick.

Similarly, Avogadro's number, $N_A$, is formally defined to be the number of carbon-12 atoms in 12 grams (0.012 kg) of unbound carbon-12 in its rest-energy electronic state. This definition relies on a precise definition of a gram, which for the past 117 years has been taken to be 1/1000$^{th}$ of the mass of "Le Gran K", a single precious platinum-iridium cylinder in Sevres, France. Various experimental approximations have resulted in the current best estimate of this number, $(6.0221415 \pm 0.0000010) \times 10^{23}$. But, since the definition of $N_A$ depends on the artifact cylinder, there has been an effort to define $N_A$ via various other methods, and hence, at least theoretically, allowing replacement of the kilogram artifact.

http://physics.nist.gov:80/News/TechBeat/9501beat.html

An alternative, purely mathematical solution to the approximation of $N_A$ is to follow the precedent set in defining the speed of light, and simply to *define* $N_A$ once and for all. Since its formal definition suggests that it counts something, it is natural to require that the definition of $N_A$ be an integer, and one that it accurate to within the present state of the art in accuracy. Here is our proposed value, $N_A^*$, of Avogadro's number:

$$N_A^* = (84446888)^3$$

It is easy to check that $N_A^*$ is within (in fact nearly dead center) the NIST standard and relative standard uncertainty of $N_A$

http://physics.nist.gov/cgi-bin/cuu/Value?na|search_for=avogadro%27s+number

so it is *exactly correct* by current standards, as was the case when a fixed value for the speed of light was selected.

Moreover, defining Avogadro's number as $N_A$* has a number of advantages over experimentally determined values:

1. $N_A$* does not vary in time, as any definition depending on an artifact will. (The long-term instability of the French artifact is well known. In fact, Le Gran K is decreasing in time, since each cleaning presumably removes atoms, and none are replaced, so formally speaking, $N_A$ is decreasing in time.)

2. $N_A$* is tractable (8 digits) and easy to remember (in contrast to the defined speed of light, for which there are mnemonics ("Since a nine-digit sequence is a bit hard to remember, there are several useful mnemonics for *c* in m/s, which use the letters on a telephone keypad: **C**onstant **W**hich **W**e **R**emember **W**ell **B**ecause **I**t's **L**ight's **V**elocity; and **A** **W**ay **W**e **R**emember **W**hat **C**onstant **I**s **L**ight's **V**elocity").

3. $N_A$* is a perfect cube, divisible by $2^9$. That it is divisible by $2^9$ simplifies many calculations, and that it is a perfect cube is consistent with Avogadro's hypothesis that equal volumes of gases with the same pressure and temperature contain the same number of molecules. Since the shape of a volume certainly affects the numbers of molecules it can contain – extremely long thin cylinders may contain *none* – it seems natural to require the shape to be a perfect geometric cube, with integral numbers of molecules along each edge, implying that the number of molecules it contains is a perfect cube.

4. $N_A$* allows acceptable experimental determination of the value of 1kg, thus offering a possible replacement for the kilogram standard. As accuracy of measuring equipment improves, the $N_A$*-derived value of a kilogram will be within NIST standards, if it is not already so. Here are the analogs of two classical experiments designed to determine $N_A$, but which may equally well be used, via $N_A$*, to determine 1kg. (Of course theoretically, a gram would now be, by *definition*, 1/12 the mass of a cube of carbon-12 atoms exactly 84446888 atoms on a side – and to within today's accuracy, that same value is also exact using $N_A$. In time, a direct measurement of the mass of the perfect $N_A$* carbon-12 cube should be possible.)

      A. Oleic acid method. (high school physics laboratory experiment).

In short, the experimenter weighs out a small amount of oleic acid, forms it into a liquid monolayer of oleic acid one molecule thick, measures its area (diameter), and using the (for high school purposes known) value for the diameter of an oleic acid molecule, calculates the number of atoms in the thin disc monolayer. Using the (again assumed known) value for the molecular weight of oleic acid, $N_A$* then yields the mass in grams of the oleic acid monolayer (which may then be used to calibrate the scales it was originally weighed on, i.e., to determine the value of 1g). For more details:

http://www.gpc.edu/~ddonald/chemlab/oleicavagno.html

B. X-ray diffraction in crystal silicon lattices. (Current state of the art).

In short, using an Si artifact sphere of diameter about 10cm of very high purity (grown using the Float Zone process, and nitrogen doped to reduce the content of swirl defects – see photos of such spheres on following websites), three quantities are measured: the volume occupied by a single Si atom (via precise knowledge of the lattice structure, taking into account impurities and self-point crystal defects, and using a scanning X-ray interferometer; the volume of the sphere (via measuring the diameter using optical interferometry) ; and the molar value of the lattice (i.e., its isotropic composition) via fluorinating the Si and measuring isotope ratios using a mass spectrometer. Knowing the volume of a single Si atom and the volume of the 10-cm artifact determines the number of atoms in the artifact. Using the determined molecular weight of the artifact, $N_A$* then yields its mass, i.e., yielding the value of 1kg.

The following two sites have detailed descriptions of the experiment.

http://www.aist.go.jp/aist_e/latest_research/2004/20040129/20040129.html

http://www.npl.co.uk/environment/pdf/avogadro.pdf

The proposal to use $N_A$* has one more marked advantage in reducing experimental errors. Using today's methods for determining Avogadro's number requires TWO distinct experiments, usually far apart in time and space, namely: first calibrating the scales (at the laboratory in US or Japan etc) with Le Gran K in France; and second, running the $N_A$ experiment. The resulting best current approximation to $N_A$ thus compounds the errors from BOTH experiments. Our proposal will not necessitate any new apparatus; rather, it eliminates the need for Le Gran K in France altogether. Precisely the same experiment that is currently used to determine $N_A$, would now simply measure 1 gram.

Ron.fox@physics.gatech.edu   School of Physics
hill@math.gatech.edu School of Mathematics

Georgia Institute of Technology
Atlanta GA 30332 USA

ADDENDUM to: A Proposed Exact Integer Value for Avogadro's Number
By Ronald F. Fox and Theodore P. Hill
April 28, 2007

Abstract:

The authors have proposed fixing Avogadro's number, $N_A$, as the exact cube of an integer. They identified the ten cubes within the NIST range of accepted values for $N_A$, any of which would suffice to redefine the kilogram consistently with current accepted values for the fundamental physical constants, and expressed a preference for the (element-independent) cube closest to the center of the NIST range. Chemists' suggest fixing a value for $N_A$ that is exactly divisible by 12 (which the closest-to-center cube is not) since if $N_A$ were divisible by 12, that would imply that one gram would be an exact integral number of carbon-12 atoms. In light of this suggestion, the authors recommend adopting the cube in that list that is two atoms shorter on each side, namely $84446886^3$, as the fixed value for $N_A$.

Article:

In [1] and [2] the authors proposed fixing Avogadro's number, $N_A$, as the exact cube of an integer. They identified the ten perfect cubes within the current NIST range of values for $N_A$, namely $84446884^3$, $84446885^3$, $84446886^3$, …,$84446893^3$, and pointed out that for the purpose of redefining the kilogram by fixing a value for $N_A$ that is consistent with the currently accepted values of all other fundamental constants, any one of those ten cubes would suffice. In [1] and [2], the authors expressed a preference for using the cube closest to the center of the NIST range, namely, $84446888^3$, which is independent of the choice of element (currently carbon-12) used as a base for defining the atomic mass unit *amu*.

In subsequent correspondence [3], the Chair of the Committee on Nomenclature, Terminology and Symbols of the American Chemical Society, Professor Paul J. Karol, informed the authors that his Committee has advocated [4] fixing a value for $N_A$ that is exactly divisible by 12, namely 602,214,180,000,000,000,000,000, with the implicit assumption that carbon-12 will continue to serve as the base element for the definitions of related physical constants. That integer is indeed divisible by 12, but it lacks intuitive physical significance, and more seriously, is heavily based on its decimal representation, whereas modern computers use binary or hexadecimal arithmetic. In this case that means that this particular 79-bit integer can not even be entered into a digital computer using IEEE standard double-precision arithmetic - hence exactness and accuracy in scientific calculations based on $N_A$ would be lost from the outset.

On the other hand, it is very easy to combine these two ideas – to use a perfect cube and an integer divisible by 12 -  and the authors hereby propose adopting one of the other perfect cubes listed above that is divisible by 12, namely

$$NA' = 84446886^3$$

as the fixed value for Avogadro's number. This number differs from our original proposal by only two atoms on each side, lies well within the current NIST range of values, is divisible by 12, and is easily entered in floating-point binary arithmetic (in hexadecimal form, it is simply 5088EA6 E03). If carbon-12 is retained as the base element, adoption of $N_A'$ as the fixed value for Avogadro's Number would result in clear and clean simplifications of other fundamental constants. For example, since $84446886^3/12 = 18 \times 14074481^3$ = (in prime factors) $2 \times 3^2 \times 1667^3 \times 8443^3$, using $N_A'$ would imply that

> 1 gram is the mass of exactly $18 \times 14074481^3$ carbon-12 atoms
>
> 1 *amu* is exactly $1/(18 \times 14074481^3)$ gram
>
> 1 mole of any entity (element, chemical compound, etc) is exactly $84446886^3$ of those entities.

(Another of the 10 cubes we listed, $84446892^3$ is even divisible by 12-cubed, but is not as close to the NIST central value).

As emphasized in [1] and [2], this new definition of the gram (and hence kilogram), could replace the unstable artifact *Le Gran K* as the definition of mass in the basic SI units, and these new definitions would likely help eliminate students' confusion about the current definitions of mole and *amu*.

[1] Fox, Ronald F.; Hill, Theodore P. A Proposed Exact Integer Value for Avogadro's Number, eprint arXiv:physics/0612087, December 8, 2006

   http://arxiv.org/ftp/physics/papers/0612/0612087.pdf

[2] Fox, Ronald F.; Hill, Theodore P, An Exact Value for Avogadro's Number, American Scientist March-April 2007

   http://www.americanscientist.org/AssetDetail/assetid/54773

[3] Karol, Paul J. Letter to the Editors of American Scientist, March 16, 2007, private correspondence.

[4] Karol, Paul J. *J. Chem. Educ.* **2004** *81* 800.

Ron.fox@physics.gatech.edu   School of Physics
hill@math.gatech.edu          School of Mathematics

Georgia Institute of Technology
Atlanta GA 30332 USA